\RequirePackage{fix-cm}
\documentclass[smallcondensed,natbib,oneside]{svjour3}     
\usepackage{graphicx}
\usepackage[colorlinks=true,allcolors=blue]{hyperref}%

\usepackage{amssymb}
\usepackage{etoolbox}
\usepackage{setspace}
\usepackage{url}
\usepackage{textcomp}
\usepackage{multirow}
\usepackage{pdflscape}
\usepackage{geometry}
\usepackage{hyperref}
\usepackage{booktabs, makecell, longtable}

\newcommand{\argmin}{\arg\!\min}

\newcommand{\ie}{i.e.,~}
\newcommand{\eg}{e.g.,~}
\begin{document}

\title{Deep Learning for Biomedical Image Reconstruction: A Survey
}


\author{Hanene Ben Yedder   \and 
        Ben Cardoen     \and 
        Ghassan Hamarneh
}


\institute{H. Ben Yedder \at
              \email{hbenyedd@sfu.ca}           
           \and
          B. Cardoen \at
        \email{bcardoen@sfu.ca} 
         \and
         G. Hamarneh  \at
        \email{hamarneh@sfu.ca}  \at
        School of Computing Science, Simon Fraser University, Canada \\
}


\maketitle

\begin{abstract}

Medical imaging is an invaluable resource in medicine as it enables to peer inside the human body and provides scientists and physicians with a wealth of information indispensable for understanding, modelling, diagnosis, and treatment of diseases.
Reconstruction algorithms entail transforming signals collected by acquisition hardware into interpretable images. Reconstruction is a challenging task given the ill-posed of the problem and the absence of exact analytic inverse transforms in practical cases.
While the last decades witnessed impressive advancements in terms of new modalities, improved temporal and spatial resolution, reduced cost, and wider applicability, several improvements can still be envisioned such as reducing acquisition and reconstruction time to reduce patient's exposure to radiation and discomfort while increasing clinics throughput and reconstruction accuracy. Furthermore, the deployment of biomedical imaging in handheld devices with small power requires a fine balance between accuracy and latency.
The design of fast, robust, and accurate reconstruction algorithms is a desirable, yet challenging, research goal. While the classical image reconstruction algorithms approximate the inverse function relying on expert-tuned parameters to ensure reconstruction performance, deep learning (DL) allows automatic feature extraction and real-time inference. Hence, DL presents a promising approach to image reconstruction with artifact reduction and reconstruction speed-up reported in recent works as part of a rapidly growing field. 
We review state-of-the-art image reconstruction algorithms with a focus on DL-based methods.
First, we examine common reconstruction algorithm designs, applied metrics, and datasets used in the literature. Then,
key challenges are discussed as potentially promising strategic directions for future research.
\end{abstract}

\section{Introduction}
\label{intro}
Biomedical image reconstruction translates signals acquired by a wide range of sensors into images that can be used for diagnosis and discovery of biological processes in cell and organ tissue.
Each biomedical imaging modality leverages signals in different bands of the electromagnetic spectrum, e.g. from gamma rays ( Positron emission tomography PET/SPECT)), X-rays (computed tomography (CT)), visible light (microscopy, endoscopy), infrared (thermal images), and radio-frequency (Nuclear magnetic resonance imaging (MRI)), as well as pressure sound waves (in ultrasound (US) imaging)~\citep{webb2003introduction}.
Reconstruction algorithms transform the collected signals into a 2, 3 or 4-dimensional image.

The accuracy of each reconstruction is critical for discovery and diagnosis.
Robustness to noise and generalization cross modality specifications' (\eg sampling pattern, rate, etc.) and imaging devices parameters' allow a reconstruction algorithm to be used in wider applications.
The time required for each reconstruction determines the number of subjects that can be diagnosed as well as the suitability of the technique in operating theatres and emergency situations.
The number of measurements needed for a high quality reconstruction impacts the exposure a patient or sample will have to endure.
Finally, the hardware requirements define whether a reconstruction algorithm can be used only in a dedicated facility or in portable devices thus dictating the flexibility of deployment.

\begin{figure*}[t!]
\centering
\includegraphics[width=0.99\textwidth]{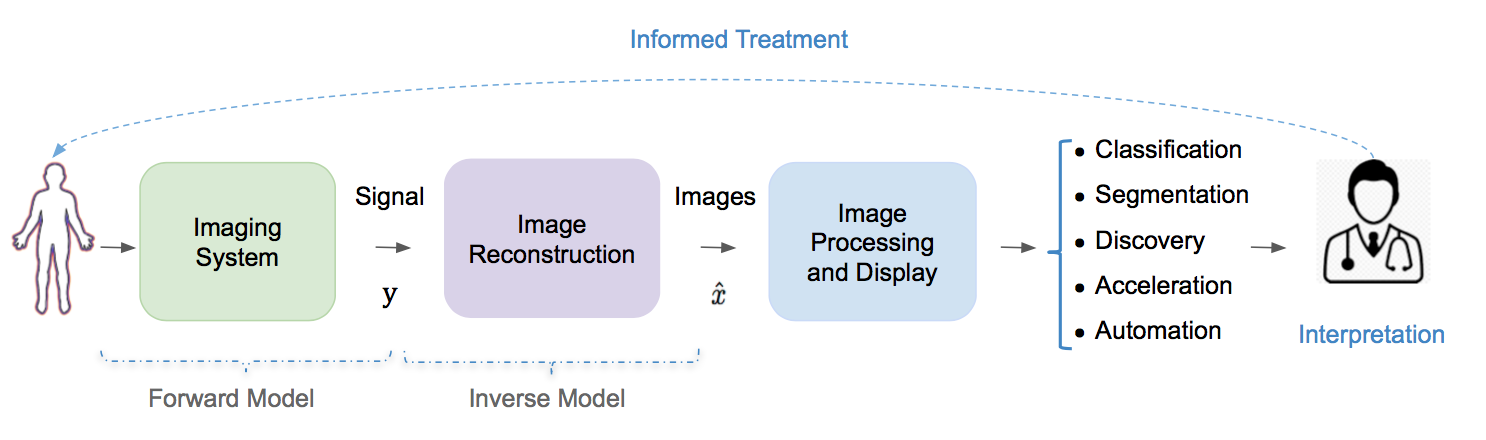}
\caption{Data flow in a medical imaging and image interpretation system. Forward model encodes the physics of the imaging system. The inverse model transforms the collected signals by the acquisition hardware into a meaningful image. The success of a diagnosis, evaluation, and treatment rely on accurate reconstruction, image visualization and processing algorithms.}
\label{Inverse}
\end{figure*}

The study of image reconstruction is an active area of research in modern applied mathematics, engineering and computer science.
It forms one of the most active interdisciplinary fields of science~\citep{fessler2017medical} given that improvement in the quality of reconstructed images offers scientists and clinicians an unprecedented insight into the biological processes underlying disease.
Fig.~\ref{Inverse} provides an illustration of the reconstruction problem and shows a typical data flow in a medical imaging system.

Over the past few years, researchers have begun to apply machine learning techniques to biomedical reconstruction to enable real-time inference and improved image quality in a clinical setting.
Here, we first provide an overview of the image reconstruction problem and outline its characteristics and challenges (Section~\ref{depth:Inverse})
and then outline the purpose, scope, and the layout of this review (Section~\ref{depth:Scope}).

\subsection{Inverse Problem and Challenges}
\label{depth:Inverse}
\subsubsection{From Output to Input}
\begin{figure*}[t!]
\centering
\includegraphics[width=0.99\textwidth]{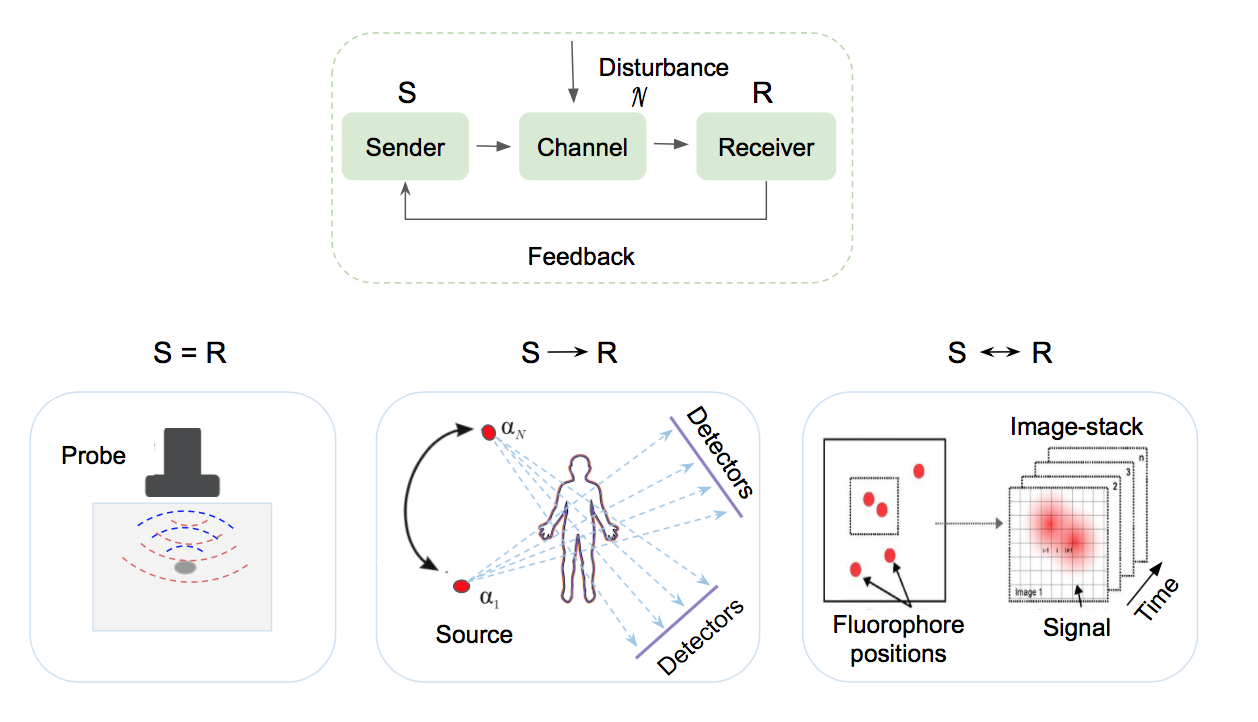}
\caption{Propagation of signals from sender to receiver. While passing
through a transmission channel, signals $s$ pick up noise (assuming additive)
along the way, until a measurement $M$ = s+$\mathcal{N}$ reaches the receiver. The properties of the received signal may, via a feedback loop, affect properties of future signal transmission. Sender and receiver modeling differ within modalities. For example the lower figure illustrates in the left ultrasound probe used to send and collect signals (S=R); in the middle: X-ray signal propagates through subject toward detector (S $\rightarrow$ R); and in the right the laser power is tuned on a feedback loop during acquisition in super resolution microscopy (S $\leftrightarrow$ R) }
\label{Biosignals}
\end{figure*}

Image reconstruction is the process of forming an interpretable image from the raw data (signals) collected by the imaging device. It is known as an inverse problem where given a set of measurements, the goal is to determine the original structure influencing the signal collected by a receiver given some signal transmission medium properties (Fig.~\ref{Biosignals}).
Let $y$
represent a set of raw acquired sensor measurements and subject to some noise $\mathcal{N}$ intrinsic to the acquisition process. The objective is to recover the spatial-domain (or spatio-temporal) image ${x} 
$ such that:
\begin{equation}
     y = \mathcal{A}(\mathcal F(x), \mathcal{N}) 
     \label{inverse}
\end{equation}
where $\mathcal F(\cdot)$ is the forward model operator that models the physics of image-formation, which can include signal propagation, attenuation, scattering, reflection and other transforms, e.g. Radon or Fourier transform. $\mathcal F(\cdot)$ can be a linear or a nonlinear operator depending on the imaging modality. $\mathcal{A}$ is an aggregation operation representing the interaction between noise and signal, in the assumption of additive noise $\mathcal{A} = +$.

While imaging systems are usually well approximated using mathematical models that define the forward model, an exact analytic inverse transform $\mathcal A^{-1}(\cdot)$ is not always possible. 
Reconstruction approaches usually resort to iteratively approximate the inverse function and often involve expert-tuned parameters and prior domain knowledge considerations to optimize reconstruction performance. 
\begin{figure}[ht!]
\centering
\includegraphics[scale=0.57, clip]{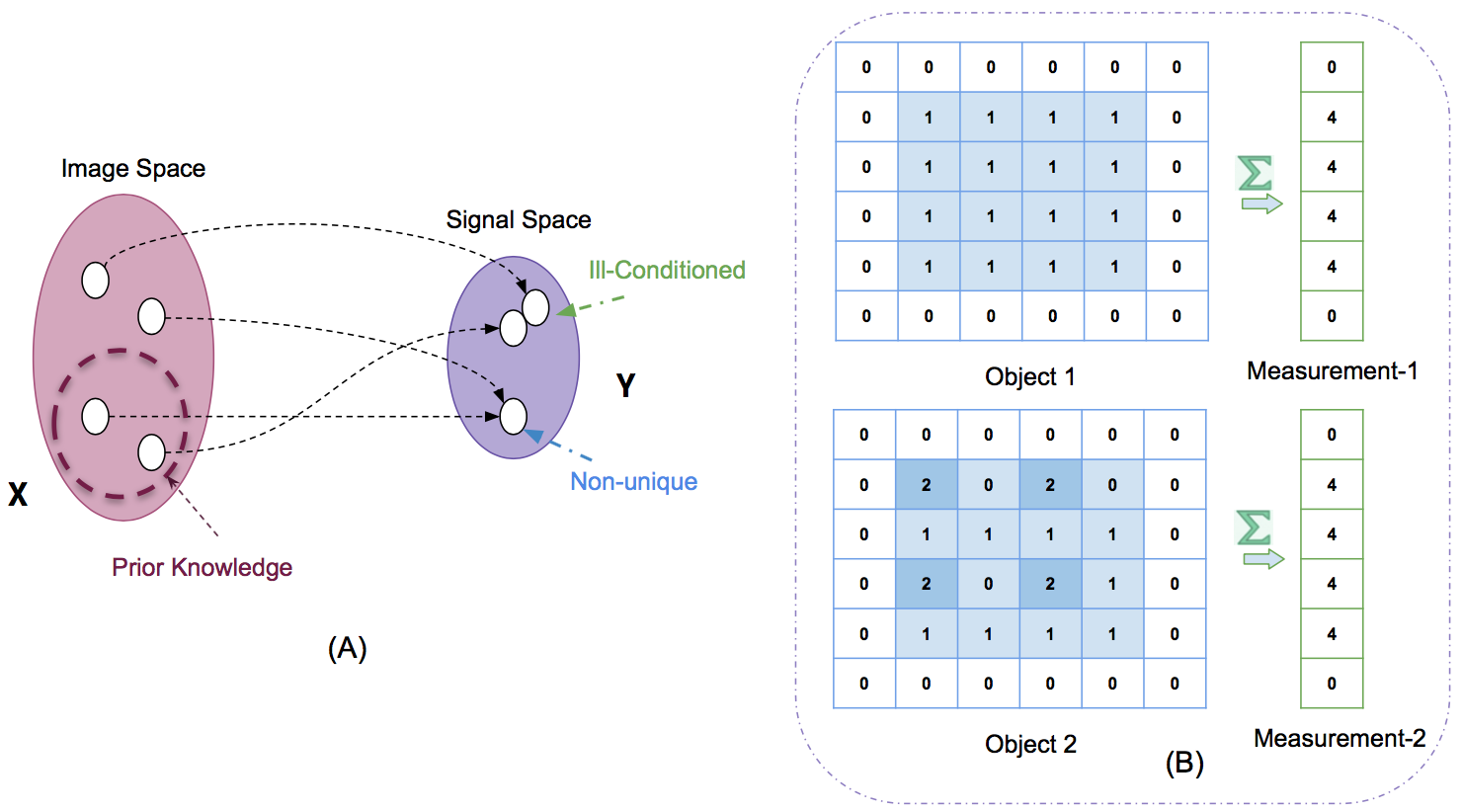}
\caption{(A) A problem is ill-conditioned when two different objects produce very close
observed signals. When the observed signals are identical and hence identical reconstructed images, inverse solution is non-unique. Prior knowledge can be leveraged to rules out certain solutions that conflict with the additional knowledge about the object beyond the measurement vectors. (B) A use case toy example of two objects with the same acquired signals. Prior knowledge about homogeneity of object rules out the second object.}


%
%
%
\label{Illposed}
\end{figure}
\subsubsection{An Ill-Posed Problem}

Image reconstruction is an ill-posed problem as there may be significantly fewer measurements $(M)$ than the number of unknowns $(N)$.
Mathematically, the problem is highly under-determined as there would be fewer equations to describe the model than unknowns ($M \ll N$) and therefore, there may be infinite consistent images that map to the same measurements (Fig.~\ref{Illposed}). 
Thus, one challenge for the reconstruction algorithm is to select the best solution among a set of potential solutions~\citep{mccann2019algorithms}. One way to reduce the solution space is to leverage domain specific knowledge by encoding priors, i.e. regularization.

Reducing data representation, such as sub-sampling in MRI or sparse-view, limited-angle data in CT, to accelerate acquisition or reduce radiation dose typically decreases the size $N$ of the measured signal $y$ while increasing its sparsity and noise level. As a consequence the ill-posedness and complexity of the reconstruction problem increase.
This brings up the need for sophisticated reconstruction algorithms with high feature extraction power to make the best use of the collected signal, capture modality-specific imaging features, and leverage prior knowledge.
Furthermore, developing high-quality reconstruction algorithms requires not only a deep understanding of both the physics of the imaging systems and the biomedical structures but also specially designed algorithms that can account for the statistical properties of the measurements and tolerate errors in the measured data. 
\subsection{Scope of this survey}
\label{depth:Scope}

The field of biomedical image reconstruction have undergone significant advances over the past few decades and can be broadly classified into two categories: conventional methods (analytical and optimization based methods) and data-driven or learning-based methods. Conventional methods (discussed in section~\ref{depth:Conventional}) are the most dominant and have been extensively studied over the last few decades with a focus on how to improve their results~\citep{vandenberghe2001iterative,jhamb2015review,assili2018review} and reduce their computational cost~\citep{despres2017review}.

\begin{figure}[ht!]
\centering
    \begin{minipage}{0.5\textwidth}
        \centering
        \includegraphics[width=0.9\textwidth]{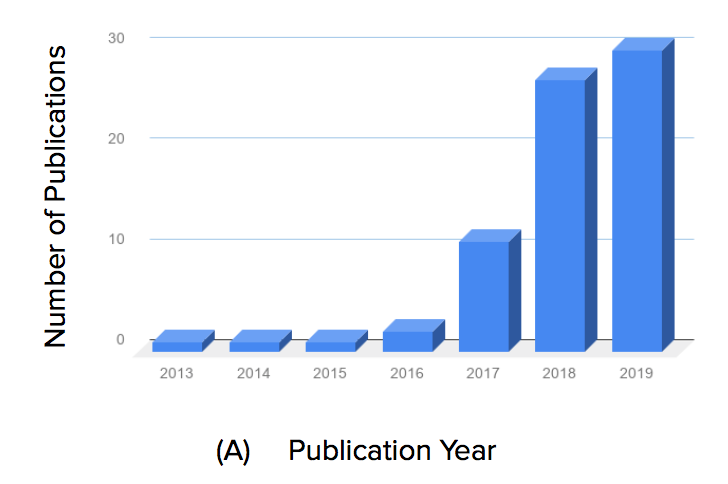} 
    \end{minipage}\hfill
    \begin{minipage}{0.5\textwidth}
        \centering
        \includegraphics[width=0.9\textwidth]{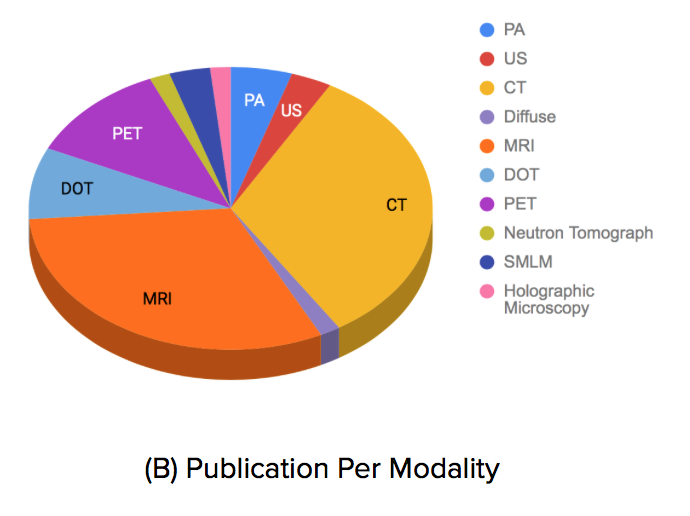} 
    \end{minipage}
\caption{(A) Number of studies, covered in this survey, on the topic of machine learning for biomedical image reconstruction versus year of publication. (B) The pie chart represents the number of studies per modality.}
\label{publication_per_year}
\end{figure}

Researchers have recently investigated deep learning (DL) approaches for various biomedical image reconstruction problems (discussed in section~\ref{depth:DLapproaches}) inspired by the success of deep learning in computer vision problems.
This topic is relatively new and has gained a lot of interest over the last few years, as shown in Fig.~\ref{publication_per_year}-A and listed in Table~\ref{Surveyed_papers}, and forms a very active field of research with numerous special journal issues~\citep{wang2016perspective,wang2018image,ravishankar2019image}. MRI and CT received the most attention among studied modalities, as illustrated in Fig.~\ref{publication_per_year}-B, given their widespread clinical usage, the availability of analytic inverse transform and the availability of public (real) datasets.

\begin{figure}[ht!]
\centering
\includegraphics[scale=0.7, clip]{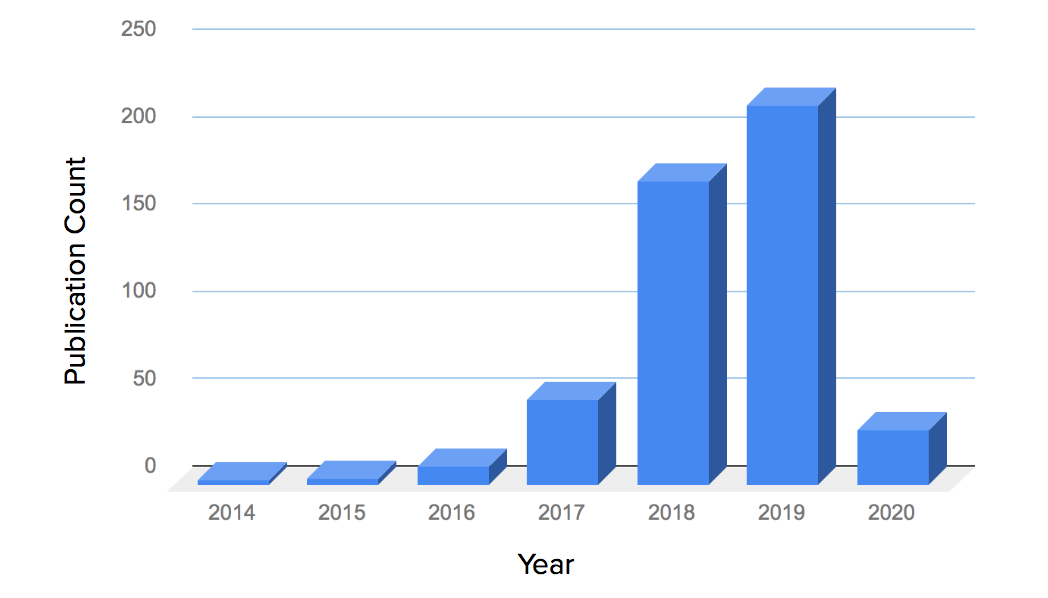}
\caption{The marked increase in publications on biomedical image reconstruction and deep learning in the past 10 years (Results obtained by PubMed query that can be found at:
\href{http://bit.ly/38ELjuz}{\url{http://bit.ly/image_recon}}).}
\label{publication_hit}
\end{figure}

To date, few surveys on machine learning-based image reconstruction approaches have been conducted. Fessler (\citeyear{fessler2017medical}) wrote a brief chronological overview on image reconstruction methods highlighting an evolution from basic analytical to data-driven models.
Recently, McCann \etal~\citeyear{mccann2019algorithms} wrote a survey on image reconstruction approaches where they presented a toolbox of operators that can be used to build imaging systems model's and showed how forward model and sparsity-based regularization can be used to solve reconstruction problems. While their review is more focused on the mathematical foundations of conventional methods, they briefly discussed data-driven approaches, their theoretical underpinning, and performance. As illustrated in Fig.~\ref{publication_hit} since its publication a great deal of work has been done warranting a review.
Similarly, Zhang \etal~(\citeyear{zhang2019review}) provided a conceptual review of some recent DL-based methods for CT with a focus on methods inspired by iterative optimization approaches and their theoretical underpinning from the perspective of representation learning and differential equations.

This survey provides an overview of biomedical image reconstruction methods with a focus on DL-based strategies, discusses their different paradigms (\eg image domain, sensor domain (raw data) or end to end learning, architecture, loss, etc. ) and how such methods can help overcome the weaknesses of conventional non-learning methods. The theoretical foundation was not emphasized in this work as it was well investigated in the aforementioned surveys. Common evaluation metrics and training dataset challenges are also discussed. 

The remainder of this paper is organized as follows: in Section~\ref{depth:Conventional} we give an overview of conventional methods discussing their advantages and limitations. 
We then introduce the key machine learning paradigms and how they are being adapted in this field complementing and improving on conventional methods.
A review of available data-sets and performance metrics is detailed in Section~\ref{depth:DLapproaches}.
Finally, we conclude by summarizing the current state-of-the-art and outlining strategic future research directions (Section~\ref{depth:Conclusion}).

\section{Conventional Image Reconstruction Approaches}
\label{depth:Conventional}

A wide variety of reconstruction algorithms have been proposed during the past few decades, having evolved from analytical methods to iterative or optimization-based methods that account for the statistical properties of the measurements and noise as well as the hardware of the imaging system~\citep{fessler2017medical}. While these methods have resulted in significant improvements in reconstruction accuracy and artifact reduction, are in routine clinical use currently, they still present some weaknesses.
A brief overview of these methods' principles is presented in this section outlining their weaknesses. 
\subsection{Analytical Methods}
Analytical methods are based on a continuous representation of the reconstruction problem and use simple mathematical models for the imaging system. 
Classical examples are the inverse of the Radon transform such as filtered back-projection (FBP) for CT and the inverse Fourier transform (IFT) for MRI. These methods are usually computationally inexpensive (in the order of ms) and can generate good image quality in the absence of noise and under the assumption of full sampling/all angles projection~\citep{mccann2019algorithms}. They typically consider only the geometry and sampling properties of the imaging system while ignoring the details of the system physics and measurement noise~\citep{fessler2017medical}. 

When dealing with noisy or incomplete measured data~\eg reducing the measurement sampling rate, analytical methods results are highly deteriorated as the signal becomes weaker. Thus, the quality of the produced image is compromised.
Thaler \etal~(\citeyear{thaler2018sparse}) provided examples of a CT image reconstruction using the FBP method for different limited projection angles and showed that analytical methods are unable to recover the loss in the signal, resulting in compromised diagnostic performance.
\subsection{Iterative Methods}
Iterative reconstruction methods, based on more sophisticated models for the imaging system's physics, sensors and noise statistics, have attracted a lot of attention over the past few decades.
They combine the statistical properties of data in the sensor domain (raw measurement data), prior information in the image domain, and sometimes parameters of the imaging system into their objective function~\citep{mccann2019algorithms}. 
Compared to analytical methods iterative reconstruction algorithms offer a more flexible reconstruction framework and better robustness to noise and incomplete data representation problems at the cost of increased computation ~\citep{ravishankar2019image}.

\begin{figure*}[ht!]
\centering
\includegraphics[ clip,width=\textwidth]{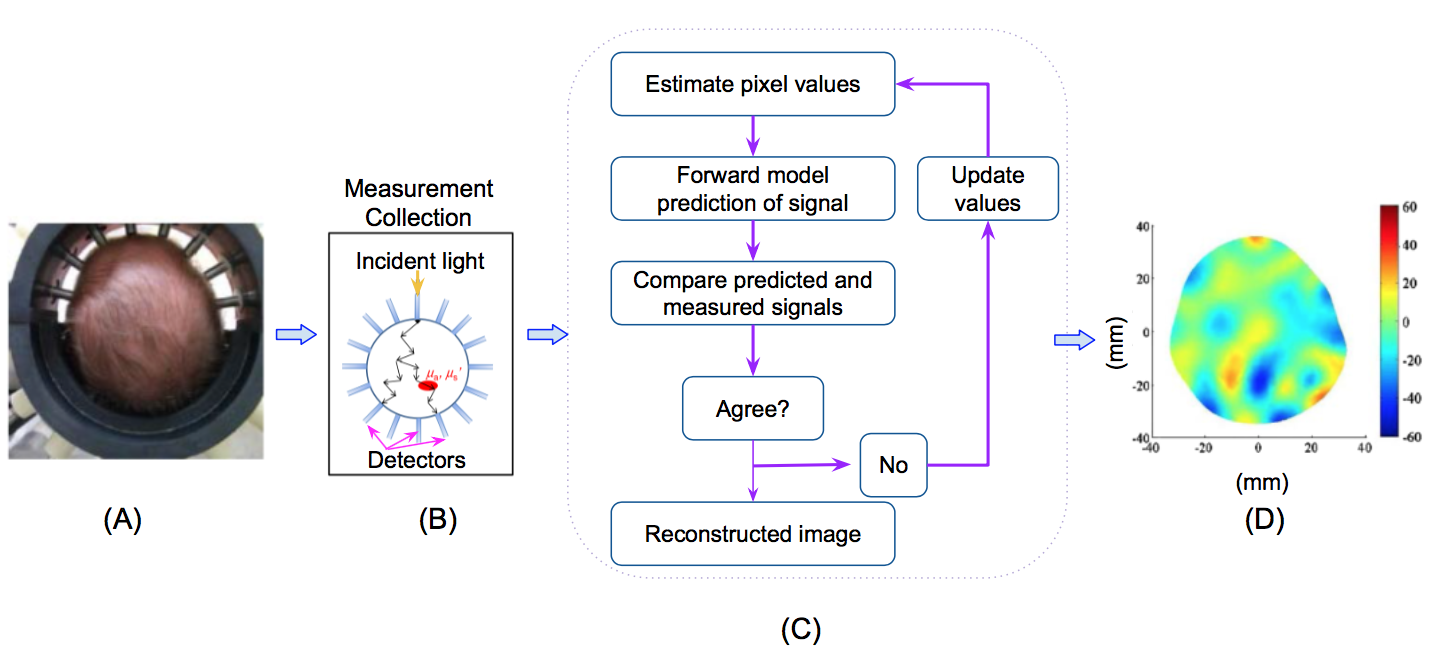}
\caption{Iterative image reconstruction workflow example. (A) Diffuse optical tomography (DOT) fibers brain probe consisting of a set of fibers for illumination and outer bundle fibers for detection. (B) Probe scheme and light propagation modelling in the head, used by the forward model. (C) Iterative approach pipeline. (D) DOT reconstructed image shows the total Hemoglobin (Hb) concentrations in the brain. (Figure licensed under CC-BY 4.0 Creative Commons license.)}
\label{Iterativeapproach}
\end{figure*}
Iterative reconstruction methods involve minimizing an objective function that usually consists of a data term and a regularization terms imposing some prior:
\begin{equation}
 \hat{x}^* = \argmin_{\hat{x}} \|\mathcal F(\hat{x}) -y\|+ \lambda \mathcal R(\hat{x})
 \label{iterative}
\end{equation}
where $||\mathcal F(\hat{x}) -y\|$ is a data fidelity term that measures the consistency of the approximate solution to the measured signal y,
$\mathcal R(\cdot)$ is a regularization term encoding the prior information about the data, and $\lambda$ is a hyper-parameter that controls the contribution of the regularization term. 
The reconstruction error is minimized iteratively until convergence.
The regularization term is often the most important part of the modeling and what researchers have mostly focused on in the literature as it vastly reduces the solution space by accounting for assumptions based on the underlying data (\eg smoothness, sparsity, spatio-temporal redundancy). 
The interested reader can refer to~\citep{dong2015image,mccann2019algorithms} for more details on regularization modeling. 
Fig.~\ref{Iterativeapproach} shows an example of an iterative approach workflow for diffuse optical tomography (DOT) imaging. 

Several image priors were formulated as sparse transforms to deal with incomplete data issues.
The sparsity idea, representing a high dimensional image $x$ by only a small number of nonzero coefficients, is one dominant paradigm that has been shown to drastically improve the reconstruction quality especially when the number of measurements $N$ or theirs signal to noise ratio (SNR) is low. 
Given the assumption that an image can be represented with a few nonzero coefficients (instead of its number of pixels), it is possible to recover it from a much smaller number of measurements.
A popular choice for a sparsifying transform is total variation (TV) that is widely studied in academic literature. The interested reader is referred to Rodriguez \etal~\citep{rodriguez2013total} for TV based algorithms modeling details.
While TV imposes a strong assumption on the gradient sparsity via the non-smooth absolute value that is more suited to piece-wise constant images, TV tends to cause artifacts such as blurred details and undesirable patchy texture in the final reconstructions. 
Recent work aimed at exploiting richer feature knowledge overcomes TV's weaknesses, for example TV-type variants~\citep{zhang2016statistical}, non-local means (NLM)~\citep{zhang2016spectral}, wavelet approaches~\citep{gao2011multi}, and dictionary learning~\citep{xu2012low}.
Non-local means filtering methods, widely used for CT~\citep{zhang2017applications}, are operational in the image domain and allow the estimation of the noise component based on multiple patches extracted at different locations in the image~\citep{sun2016deep}.

Overall, although iterative reconstruction methods showed substantial accuracy improvements and artifact reductions over the analytical ones, they still face three major weaknesses: First, iterative reconstruction techniques tend to be vendor-specific since the details of the scanner geometry and correction steps are not always available to users and other vendors. Second, there are substantial computational overhead costs associated with popular iterative reconstruction techniques due to the load of the projection and back-projection operations required at each iteration. 
The computational cost of these methods is often several orders of magnitude higher than analytical methods, even when using highly-optimized implementations. A trade-off between real-time performance and quality is made in favor of quality in iterative reconstruction methods due to their non-linear complexity of quality in function of the processing time.
Finally, the reconstruction quality is highly dependent on the regularization function form and the related hyper-parameters settings as they are problem-specific and require non-trivial manual adjustments. Over-imposing sparsity ($\mathcal L1$ penalties) for instance can lead to cartoon-like artifacts. Proposing a robust iterative algorithm is still an active research area~\citep{sun2019online,moslemi2020estimation}.
\begin{figure}[t!]
\centering
\includegraphics[scale=0.55,clip]{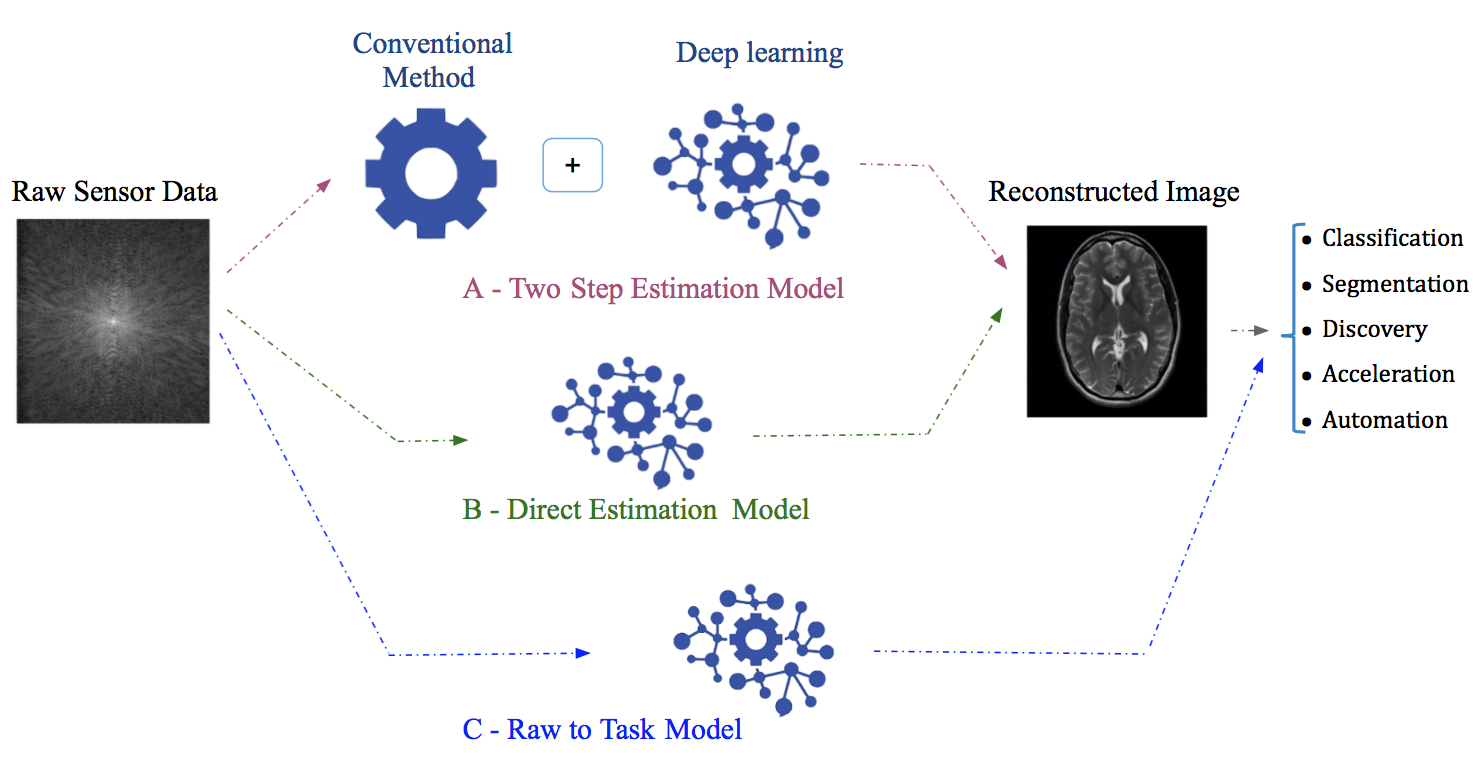}
\caption{Visualization of common deep learning-based reconstruction paradigms from raw sensor data. (A) A two-step processing model is shown where deep learning complements conventional methods (Section \ref{depth:Two steps estimation model}). A typical example would be to pre-process the raw sensor data using a conventional approach and enhance the resulting image with a deep learning model or vise versa. (B) An end-to-end model is shown: The image is directly estimated from the raw sensor data with a deep learning model (Section \ref{Direct Estimation Model}).(C) Task results can be directly inferred with or without explicit image reconstruction (Section \ref{depth:Raw to task}).} 
\label{DLdirect/pre}
\end{figure}
\section{Deep Learning Based Image Reconstruction }
\label{depth:DLapproaches}
To further advance biomedical image reconstruction, a more recent trend is to exploit deep learning techniques for solving the inverse problem to improve resolution accuracy and speed-up reconstruction results.
As a deep neural network represents a complex mapping, it can detect and leverage features in the input space and build increasingly abstract representations useful for the end-goal. Therefore, it can better exploit the measured signals by extracting more contextual information for reconstruction purposes.
In this section, we summarize works using DL for inverse problems in imaging.

Learning-based image reconstruction is driven by data where a training dataset is used to tune a parametric reconstruction algorithm that approximates the solution to the inverse problem with a significant one-time, offline training cost that is offset by a fast inference time. There is a variety of these algorithms, with some being closely related to conventional methods and others not.
While some methods considered machine learning as a reconstruction step by combining a traditional model with deep modeling to recover the missing details in the input signal or enhance the resulting image (Section~\ref{depth:Two steps estimation model}, some others considered a more elegant solution to reconstruct an image from its equivalent initial measurements directly by learning all the parameters of a deep neural network, in an end-to-end fashion, and therefore approximating the underlying physics of the inverse problem (Section~\ref{Direct Estimation Model}), or even going further and solving for the target task directly (Section~\ref{depth:Raw to task}). Fig.~\ref{DLdirect/pre} shows a generic example of the workflow of these approaches. Table~\ref {Surveyed_papers} surveys various papers based on these different paradigms and provides a comparison in terms of used data (Table~\ref {Surveyed_papers}-Column "Mod.","Samp.","D"), architecture (Table~\ref {Surveyed_papers}-Column "TA","Arch."), loss and regularization (Table~\ref {Surveyed_papers}-Column "Loss","Reg."), augmentation (Table~\ref {Surveyed_papers}-Column "Aug."), etc .

\subsection{Deep Learning as Processing Step: Two Step Image Reconstruction Models}
\label{depth:Two steps estimation model}
Complementing a conventional image reconstruction approach with a DL-model enables improved accuracy while reducing the computational cost. The problem can be addressed either in the sensor domain (pre-processing) or the image domain (post-processing) (Fig.~\ref{DLdirect/pre}-A, Table~\ref {Surveyed_papers}-Column "E2E").

\subsubsection{A Pre-Processing Step (Sensor Domain)}
\label{depth:Pre-Processing Step}

The problem is formulated as a regression in the sensor domain from incomplete data representation (\eg sub-sampling, limited view, low dose) to complete data (full dose or view) using DL methods and has led to enhanced results~\citep{hyun2018deep,liang2018improve,cheng2018highly}. 
The main goal is to estimate, using a DL model, missing parts of the signal that have not been acquired during the acquisition phase in order to input a better signal to an analytical approach for further reconstruction. 

Hyun~\etal (\citeyear{hyun2018deep}) proposed a k-space correction based U-Net to recover the unsampled data followed by an IFT to obtain the final output image. They demonstrated artifact reduction and morphological information preservation when only 30\% of the k-space data is acquired.
Similarly, Liang \etal~(\citeyear{liang2018improve}) proposed a CT angular resolution recovery based on deep residual convolutional neural networks (CNNs) for accurate full view estimation from unmeasured views while reconstructing images using filtered back projection. Reconstruction demonstrated speed-up with fewer streaking artifacts along with the retrieval of additional important image details.
Unfortunately, since noise is not only present in the incomplete data acquisition case, but also in the full data as well, minimizing the error between the reference and the predicted values can cause the model to learn to predict the mean of these values. As a result, the reconstructed images can suffer from lack of texture detail.

Huang~\etal~(\citeyear{huang2019data}) argue that DL-based methods can fail to generalize to new test instances given the limited training dataset and DL's vulnerability to small perturbations especially in noisy and incomplete data case. 
By constraining the reconstructed images to be consistent with the measured projection data, while the unmeasured information is complemented by learning based methods, reconstruction quality can be improved.
DL predicted images are used as prior images to provide information in missing angular ranges first followed by a conventional reconstruction algorithm to integrate the prior information in the missing angular ranges and constrain the reconstruction images to be consistent to the measured data in the acquired angular range.

Signal regression in the sensor domain reduces signal loss enabling improved downstream results from the coupled analytic method.
However, the features extracted by DL methods are limited to the sensor domain only while analytical methods' weaknesses are still present in afterword processing. 

\subsubsection{A Post-Processing Step (Image Domain)}
The regression task is to learn the mapping between the low-quality reconstructed image and its high-quality counterpart.
Although existing iterative reconstruction methods improved the reconstructed image quality, they remain computationally expensive and may still present reconstruction artifacts in the presence of noise or incomplete information, e.g. sparse sampling of data.
The main difficulty arises from the non-stationary nature of the noise and serious streaking artifacts due to information loss~\citep{chen2012thoracic,al2003common}. These noise and artifacts are challenging to isolate as they may have strong magnitudes and do not obey specific model distributions in the image domain~\citep{wang2016accelerating}.
The automatic learning and detection of complex patterns offered by deep neural networks offers a clear advantage in this use case over handcrafted filters.

Given an initial reconstruction estimate from a direct inverse operator \eg FBP \citep{sun2018efficient,gupta2018cnn,chen2017low}, IFT~\citep{wang2016accelerating}, or few iterative approach steps~\citep{jin2017deep,cui2019ct,xu2017200x}, deep learning is used to refine the initialized reconstruction and produce the final reconstructed image. 
For example, Chen \etal ~(\citeyear{chen2017lowb}) used an
autoencoder to improve FBP results on a limited angle CT projection.
Similarly, Jin~\etal ~(\citeyear{jin2017deep}) enhanced a direct inversion formula extracted from a spectral iterative algorithm via subsequent filtering by a U-Net to reduce artifacts. These architectures are feasible ways to capture images structure as dimensionality reduction.

Generative adversarial networks (GAN)~\citep{goodfellow2014generative} were leveraged to improve the quality of reconstructed images. 
Wolterink~\etal~(\citeyear{wolterink2017generative}) proposed to train an adversarial network to estimate full-dose CT images from low-dose CT ones and showed empirically that an adversarial network improves the model's ability to generate images with reduced aliasing artifacts.
Interestingly, they showed that combining squared error loss with adversarial loss can lead to a noise distribution similar to that in the reference full-dose image even when no spatially aligned full-dose and low dose scans are available.

Yang~\etal~(\citeyear{yang2017dagan}) proposed a deep de-aliasing GAN (DAGAN) for compressed sensing MRI reconstruction that resulted in reduced aliasing artifacts while preserving texture and edges in the reconstruction. Remarkably, a combined loss function based on content loss ( consisting of a pixel-wise image domain loss, a frequency domain loss and a perceptual VGG loss) and adversarial loss were used. While frequency domain information was incorporated to enforce similarity in both the spatial (image) and the frequency domains, a perceptual VGG coupled to a pixel-wise loss helped preserve texture and edges in the image domain.

Combining DL and conventional methods reduce the computational cost but has its own downsides. For instance, the features extracted by DL methods are highly impacted by the results of the conventional methods, especially in case of limited measurements and the presence of noise where the initially reconstructed image may contain significant and complex artifacts that may be difficult to remove even by DL models.
In addition, the information missing from the initial reconstruction is challenging to be reliably recovered by post-processing like many inverse problems in the computer vision literature
such as image inpainting.
Furthermore, the number of iterations required to obtain a reasonable initial image estimate using an iterative method can be hard to define and generally requires a long run-time (in the order of several min) to be considered for real-time scanning. 
Therefore, the post-processing approach may be more suitable to handle initial reconstructions that are of relatively good quality and not drastically different from the high-quality one.

\subsection{End-to-End Image Reconstruction: Direct Estimation}
\label{Direct Estimation Model}
An end-to-end solution leverages the image reconstruction task directly from sensor-domain data using a deep neural network by mapping sensor measurements to image domain while approximating the underlying physics of the inverse problem. (Fig.~\ref{DLdirect/pre}-B). 
This direct estimation model may represent a better alternative as it benefits from the multiple levels of abstraction and the automatic feature extraction capabilities of deep learning models.


Given pairs of measurement vectors $y$ and their corresponding ground truth images $x$ (that produce $y$), the goal is to optimize the parameters $\theta$ of a neural network 
in an end-to-end manner to learn the mapping between the measurement vector $y$ and its reconstructed tomographic image $x$, which recovers the parameters of underlying imaged tissue. Therefore, we seek the inverse function $\mathcal A^{-1}(\cdot)$ that solves:

\begin{equation}
\setlength{\abovecaptionskip}{-10pt}
  \theta{^*} = \argmin_{\theta} \mathcal L\left(\mathcal A^{-1}(y,\theta), x \right)+ \lambda \mathcal R(\mathcal A^{-1}(y,\theta))
\end{equation}
where $\mathcal L$ is the loss function of the network that, broadly, penalizes the dissimilarity between the estimated reconstruction and the ground truth. The regularization term $\mathcal R$ is often introduced to prevent over-fitting with $\mathcal L1$ or $\mathcal L2$ norms being the most common choices. 

Recently, several paradigms have emerged for end-to-end DL-based reconstruction the most common of which are generic DL models and DL models that unroll an iterative optimization.

\subsubsection{Generic Models}
Although some proposed models rely on multilayer perceptron (MLP) feed-forward artificial neural network~\citep{pelt2013fast,boublil2015spatially,feng2018back,wang2019optimization}, CNNs remain the most popular generic reconstruction models mainly due to their data filtering and features extraction capabilities. Specifically, encoder-decoder~\citep{nehme2018deep,haggstrom2019deeppet},
U-Net~\citep{waibel2018reconstruction}, ResNet~\citep{cai2018end} and decoder like architecture~\citep{yoon2018efficient,wu2018direct,zhu2018image} are the most dominant architectures as they rely on a large number of stacked layers to enrich the level of features extraction. A set of skip connections enables the later layers to reconstruct the feature maps with both the local details and the global texture and facilitates stable training when the network is very deep.

The common building blocks of neural network architectures are convolutional layers, batch normalization layers, and rectified linear units (ReLU). ReLU is usually used to enforce information non-negativity properties, given that the resulting pixels values represent tissue properties \eg chromophores concentration maps~\citep{yoo2019}, refractive index~\citep{sun2016deep}, and more examples in Table~\ref{Surveyed_papers}. Batch normalization is used to reduce the internal covariate shift and accelerates convergence. The resulting methods can produce relatively good reconstructions in a short time and can be adapted to other modalities but require a large training dataset and good initialization parameters.
Table~\ref{Surveyed_papers}-Column "E2E" (check-marked) surveys many papers with various architecture, loss, and regularization for 2D,3D and 4D different modalities.

Zhu~\etal (\citeyear{zhu2018image}) proposed a manifold learning framework based decoder neural network to emulate the fast-Fourier transform (FFT) and learn an end-to-end mapping between k-space data and image domains where they showed artifact reduction and reconstruction speed up.
However, when trained with an $\mathcal L1$ or $\mathcal L2$ loss only, a DL-based reconstructed image still exhibits blurring and information loss, especially when used with incomplete data. Similarly, Ben Yedder~\etal~(\citeyear{yedder2018deep}) proposed a decoder like model for DOT image reconstruction. While increased reconstruction speed and lesion localization accuracy are shown, some artifacts are still present in the reconstructed image when training with $\mathcal L2$ loss only. This motivated an improved loss function in their follow-up work~\citep{yedder2019limited} where they suggested combining $\mathcal L2$ with a Jaccard loss component to reduce reconstructing false-positive pixels.

Thaler~\etal~\citep{thaler2018sparse} proposed a Wasserstein GAN (WGAN) based architecture for improving the image quality for 2D CT image slice reconstruction from a limited number of projection images using a combination of $\mathcal L1$ and adversarial losses. Similarly, \cite{ouyang2019ultra} used a GAN based architecture with a task-specific perceptual and $\mathcal L1$ losses to synthesize PET images of high quality and accurate pathological features.

To further enhance results and reduce artifacts due to motion and corruption of k-space signal, Clough~\etal~(\citeyear{oksuz2019detection}) proposed a recurrent convolutional neural network (RCNN) to reconstruct high quality dynamic cardiac MR images while automatically detecting and correcting motion-related artifacts and exploiting the temporal dependencies within the sequences. Proposed architecture included two sub-networks trained jointly: an artifact detection network that identifies potentially corrupted k-space lines and an RCNN for reconstruction.

To relax the requirement of a large number of training samples, a challenging requirement in a medical setting, simulating data was proposed as an alternative source of training data. However, creating a realistic synthetic dataset is a challenging task in itself as it requires careful modeling of the complex interplay of factors influencing real-world acquisition environment.
To bridge the gap between the real and in silico worlds, transfer learning provides a potential remedy as it helps transfer the measurements from the simulation domain to the real domain by keeping the general attenuation profile while accounting for real-world factors such as scattering, etc.

Ben Yedder~\etal (\citeyear{yedder2019limited}) proposed a supervised sensor data distribution adaptation based MLP to take advantage of cross-domain learning and reported
accuracy enhancement in detecting tissue abnormalities.
Zhou~\etal~(\citeyear{zhou2019limited}) proposed unsupervised CT sinograms adaptation,
based on CycleGAN and content consistent regularization, to further alleviate the need for real measurement-reconstruction pairs. Interestingly, the proposed method integrated the measurement adaptation network and the reconstruction network in an end-to-end network to jointly optimize the whole network.

Although several generic DL architectures and loss functions have been explored to further enhance reconstruction results in different ways (resolution, lesion localization, artifact reduction, etc.), a DL-based method inherently remains a black-box that can be hard to interpret.
Interpretability is key not only for trust and accountability in a medical setting but also to correct and improve the DL model.

\subsubsection{Unrolling Iterative Methods}

Unrolling conventional optimization algorithms into a DL model has been suggested by several works~\citep{qin2018convolutional,schlemper2017deep,wurfl2018deep,sun2016deep,adler2018learned} in order to combine the benefits of traditional iterative methods and the expressive power of deep models (Table~\ref {Surveyed_papers}-Column "E2E").
Rajagopal~\etal(\citeyear{rajagopal2019towards}) proposed a theoretical framework demonstrating how to leverage iterative methods to bootstrap network performance while preserving network convergence and interpretability featured by the conventional approaches. 

Deep ADMM-Net~\citep{sun2016deep} was the first proposed model reformulating the iterative reconstruction ADMM (alternating direction method of multipliers) algorithm into a deep network for accelerating MRI reconstruction, where each stage of the architecture corresponds to an iteration in the ADMM algorithm. 
In its iterative scheme, the ADMM algorithm requires tuning of a set of parameters that are difficult to determine adaptively for a given data set. 
By unrolling the ADMM algorithm into a deep model, the tuned parameters are now all learnable from the training data.
The ADMM-Net was later further improved to Generic-ADMM-Net~\citep{Yang2017b} where a different variable splitting strategy was adopted in the derivation of the ADMM algorithm and demonstrated state-of-the-art results with a significant margin over the BM3D-based algorithm~\citep{Dabov2007}.

Similarly, the PD-Net~\citep{adler2018learned} adopted neural networks to approximate the proximal operator by unrolling the primal-dual hybrid gradient algorithm~\citep{Chambolle2011} and demonstrated performance boost compared with FBP and handcrafted reconstruction models.

In like manner, Schlemper~\etal~(\citeyear{schlemper2017deep}) proposed a cascade convolutional network that embeds the structure of the dictionary learning-based method while allowing end-to-end parameter learning. 
The proposed model enforces data consistency in the sensor and image domain simultaneously, reducing the aliasing artifacts due to sub-sampling.
An extension for dynamic MR reconstructions~\citep{Schlemper2018} exploits the inherent redundancy of MR data.

While, the majority of the aforementioned methods used shared parameters over iterations only, Qin \etal~(\citeyear{qin2018convolutional}) proposed to propagate learnt representations across both iteration and time. Bidirectional recurrent connections over optimization iterations are used to share and propagate learned representations across all stages of the reconstruction process and learn the spatio–temporal dependencies. The proposed deep learning based iterative algorithm can benefit from information extracted at all processing stages and mine the temporal information of the sequences to improve reconstruction accuracy.
The advantages of leveraging temporal information was also demonstrated in single molecule localization microscopy~\citep{cardoen2019ergo}. An LSTM was able to learn an unbiased emission density prediction in a highly variable frame sequence of spatio-temporally separated fluorescence emissions.
In other words, joint learning over the temporal domain of each sequence and across iterations leads to improved de-aliasing.

\begin{figure*}[ht!]
\centering
\includegraphics[scale=0.6, clip,width=\textwidth]{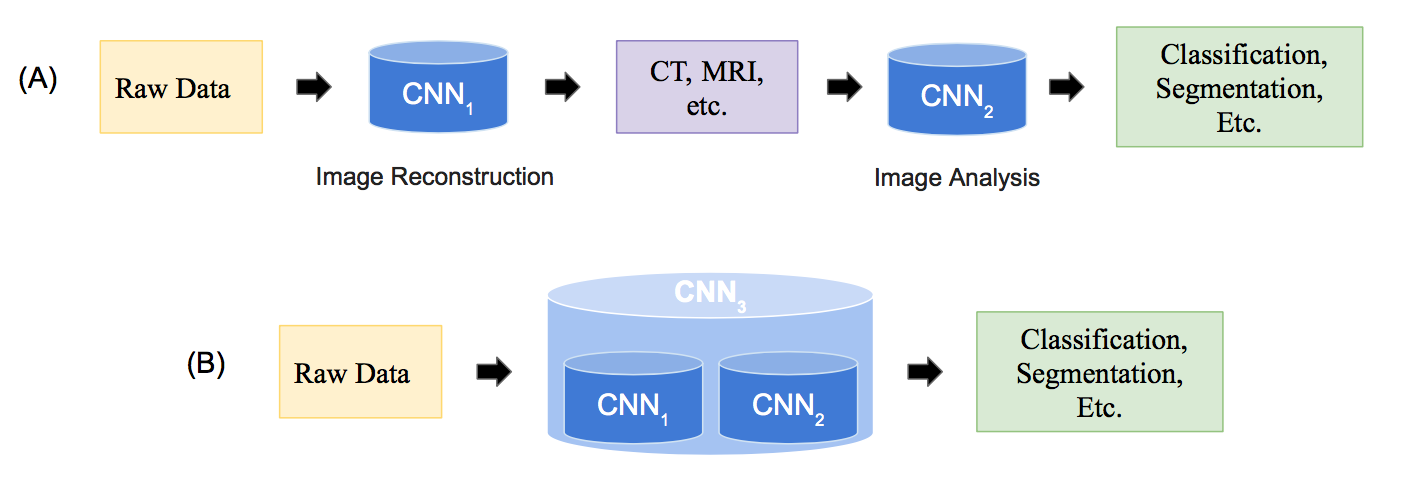}
\caption{(A) Biomedical image processing workflow usually involves two steps optimized independently (reconstruction and image analysis) for diagnosis purposes. (B) Jointly solving these tasks using a unified model allows joint parameters tuning and feature sharing.} 

\label{Joint_Model}
\end{figure*}

\subsection{Raw-to-task Methods}
\label{depth:Raw to task}
Typical data flow in biomedical imaging involves solving the image reconstruction task followed by preforming an image processing task (\eg segmentation, classification) (Fig~\ref{Joint_Model}-A). Although each of these two tasks (reconstruction and image processing) is solved separately, the useful clinical information extracted for diagnosis by the second task is highly dependent on the reconstruction results. In the raw-to-task paradigm, task results are directly inferred from raw data, where image reconstruction and processing are lumped together and reconstructed image may not be necessarily outputted (Fig.~\ref{DLdirect/pre}-C, Fig~\ref{Joint_Model}-B) 

Jointly solving for different tasks using a unified model is frequently considered in the computer vision field, especially for image restoration~\citep{Sbalzarini2016}, and has lead to improved results than solving tasks sequentially~\citep{Paul2013}.
The advantages explain the recent attention this approach received in biomedical image reconstruction~\citep{Sun2019, huang2019fr}.
For instance, a unified framework allows joint parameters tuning for both tasks and feature sharing where the two problems regularize each other when considered jointly. In addition, when mapping is performed directly from the sensor domain, the joint task can even leverage sensor domain features for further results enhancing while it can be regarded as a task-based image quality metric that is learned from the data.
Furthermore, Sbalzarini (\citeyear{Sbalzarini2016}) argues that solving ill-posed problems in sequence can be more efficient than in isolation. 
Since the output space of a method solving an inverse problem is constrained by forming the input space of the next method, the overall solution space contracts. Computational resources can then be focused on this more limited space improving runtime, solution quality or both.

\section{Medical Training Datasets}
\label{depth:Datasets}
The performance of learning-based methods is dictated to a large extent by the size and diversity of the training dataset.
In a biomedical setting, the need for large, diverse, and generic datasets is non-trivial to satisfy given constraints such as patient privacy, access to acquisition equipment and the problem of divesting medical practitioners to annotate accurately the existing data.
In this section, we will discuss how researchers address the trade-offs in this dilemma and survey the various publicly available dataset type used in biomedical image reconstruction literature. 

Table~\ref {Surveyed_papers}-Columns "Data", "Site" and "Size" summarise details about dataset used by different surveyed papers, which are broadly classified into clinical (real patient), physical phantoms, and simulated data. The sources of used datasets have been marked in the last column of Table~\ref {Surveyed_papers}-Columns "Pub". Data" in case of their public availability to other researchers. Since phantom data are not commonly made publicly available, the focus was mainly given to real and simulated data whether they are publicly available or as part of challenges. Used augmentation techniques have been mentioned in Table~\ref {Surveyed_papers}-Columns "Aug". Remarkably, augmentation is not always possible in image reconstruction task especially in sensor domain given the non-symmetries of measurements in some case, the nonlinear relationship between raw and image data, and the presence of other phenomena(\eg scattering).
We herein survey the most common source of data and discuss their pros and cons.
\subsection{Real-World Datasets}
Some online platforms (\eg~\citeauthor{giveascan}, ~\citeauthor{mridata}, MGH-USC HCP (\citeyear{MGH-USCHCP}), and ~\citeauthor{ukbiobank}) made the initiative to share datasets between researchers for image reconstruction task. ~\citeauthor{mridata}, for example, is an open platform for sharing clinical MRI raw k-space datasets. The dataset is sourced from acquisitions of different manufacturers, allowing researchers to test the sensitivity of their methods to overfitting on a single machine's output while may require the application of transfer-learning techniques to handle different distributions. 
As of writing, only a subset of organs for well known modalities \eg MRI and CT are included (Table~\ref {Surveyed_papers}-Columns "Site").
Representing the best reconstruction images acquired for a specific modality, the pairs of signal-image form a gold standard for reconstruction algorithms.
Releasing such data, while extremely valuable for researchers, is a non-trivial endeavour where legal and privacy concerns have to be taken into account by, for example, de-anonymization of the data to make sure no single patient or ethnographically distinct subset of patients can ever be identified in the dataset. Source of
real-word used datasets on surveyed papers has been marked in Table~\ref {Surveyed_papers}-Columns "Pub. Data" where they sizes remain relatively limited to allow a good generalization of DL-based methods.

\subsection{Physics-Based Simulation}
Physics-based simulation~\citep{Schweiger2014,Harrison2010,haggstrom2016dynamic} provides an alternative source of training data that allows generating a large and diversified dataset. 
The accuracy of a physical simulation with respect to real-world acquisitions increases at the cost of an often super-linear increase in computational resources.
In addition, creating realistic synthetic datasets is a nontrivial task as it requires careful modeling of the complex interplay of factors influencing real-world acquisition environment. 
With a complete model of the acquisition far beyond computational resources, a practitioner needs to determine how close to reality the simulation needs to be in order to allow the method under development to work effectively.
Transfer learning provides a potential remedy to bridge the gap between real and in silico worlds and alleviates the need for a large clinical dataset~\citep{zhu2018image,yedder2019limited}.
In contrast, the approach of not aiming for complete realism but rather using the simulation as a tool to sharpen the research question can be appropriate.
Simulation is a designed rather than a learned model. For both overfitting to available data is undesirable.
The assumptions underlying the design of the simulation are more easily verified or shown not to hold if the simulation is not fit to the data, but represents a contrasting view.
For example, simulation allows the recreation and isolation of edge cases where a current approach is performing sub-par. As such simulation is a key tool for hypothesis testing and validation of methods during development.
For DL-based methods the key advantage simulation offers is the almost unlimited volume of data that can be generated to augment limited real-world data in training. With the size of datasets as one of the keys determining factors for DL-based methods leveraging simulation is essential.
Surveyed papers that used simulated data as a training or augmentation data have been marked in Table~\ref {Surveyed_papers}-Columns "Pub".

\subsection{Challenge Dataset}
There are only a few challenge (competition) datasets for image reconstruction task \eg LowDoseCT~(\citeyear{LowDoseCT2016}) FastMRI~\citep{zbontar2018fastmri}, Fresnel~\citep{geffrin2005free} and SMLM challenge~\citep{holden2016imaging} that includes raw measurements. Simulating incomplete signals by degrading the high-quality acquired signals while keeping their corresponding high-quality images pair was also explored.
Alternatively, researchers collect high-quality images from other medical imaging challenges, \eg segmentation (MRBrainS challenge~\citep{mendrik2015mrbrains}, \cite{MICCAIchallenge2013}), and use simulation, using a well known forward model, to generate full and/or incomplete sensor domain pairs. 
Here again, only a subset of body scans and diseases for well-studied modalities are publicly available as highlighted in Table~\ref {Surveyed_papers}-Columns "Site" and "Pub. Data" .


\section{Reconstruction Evaluation Metrics}
\label{depth:Metrics}
\subsection{Quality}
Measuring the performance of the reconstruction approaches is usually performed using three metrics, commonly applied in computer vision, in order to access the quality of the reconstructed images. These metrics are the root mean squared error (RMSE) or normalized mean squared error, structural similarity index (SSIM) or its extended version multiscale structural similarity \citep{wang2003multiscale}, and peak signal to noise ratio (PSNR).

While RMSE measures the pixel-wise intensity difference between ground truth and reconstructed images by evaluating pixels independently, ignoring the overall image structure,
SSIM, a perceptual metric, quantifies visually perceived image quality and captures structural distortion. SSIM combines luminance, contrast, and image structure measurements and ranges between [0,1] where the higher SSIM value the better and SSIM = 1 means that the two compared images are identical.


PSNR (Eq.~\ref{eqn:PSNR}) is a well-known metric for image quality assessment which provides similar information as the RMSE but in units of dB while measuring the degree to which image information rises above background noise. Higher values of PSNR indicate a better reconstruction.
\begin{equation}
PSNR = 20 . {\log_{10} { \left( \frac{X_{max}}{ \sqrt{MSE}} \right)}}
\label{eqn:PSNR}
\end{equation}

\noindent where $X_{max}$ is the maximum pixel value of the ground truth image. 

Illustrating modality specific reconstruction quality is done by less frequently used metrics such as contrast to noise ratio (CNR)~\citep{wu2018direct} for US.
Normalized mutual information (NMI) is a metric used to determine the mutual information shared between two variables, in this context image ground truth and reconstruction~\citep{wu2018direct,zhou2019limited}. When there is no shared information NMI is 0, whereas if both are identical a score of 1 is obtained.
To illustrate NMI's value, consider two images X, Y with random values. When generated from two different random sources, X and Y are independent, yet RMSE(X,Y) can be quite small.
When the loss function minimizes RMSE, such cases can induce stalled convergence at a suboptimal solution due to a constant gradient. NMI, on the other hand, would return zero (or a very small value), as expected.
The intersection over union, or Jaccard index, is leveraged to ensure detailed accurate reconstruction~\citep{yedder2018deep,Sun2019}. In cases where the object of interest is of variable size and small with respect to the background, an RMSE score is biased by matching the background rather than the target of interest. In a medical context, it is often small deviations (e.g. tumors, lesions, deformations) that are critical to diagnosis. Thus, unlike computer vision problems where little texture changes might not alter the overall observer's satisfaction, in medical reconstruction, further care should be taken to ensure that practitioners are not misled by a plausible but not necessarily correct reconstruction. 
Care should be taken to always adjust metrics with respect to their expected values under an appropriate random model~\citep{gates2017impact}. 
The understanding of how a metric responds to its input should be a guideline to its use. As one example, the normalization method in NMI has as of writing no less than 6~\citep{gates2017impact} alternatives with varying effect on the metric.
Table~\ref{Surveyed_papers}-Columns "Metrics" surveys the most frequently used metrics on surveyed papers.

\subsection{Inference Speed}
With reconstruction algorithms constituting a key component in time-critical diagnosis or intervention settings, the time complexity is an important metric in selecting methods.
Two performance criteria are important in the context of time: Throughput measures how many problem instances can be solved over a time period, and latency measures the time needed to process a single problem instance. In a non-urgent medical setting, a diagnosing facility will value throughput more than latency. In an emergency setting where even small delays can be lethal, latency is critical above all.
For example, if a reconstruction algorithm is deployed on a single device it is not unexpected for there to be waiting times for processing. As a result latency, if the waiting time is included, will be high and variable, while throughput is constant.
In an emergency setting there are limits as to how many devices can be deployed, computing results on scale in a private cloud on the other hand can have high throughput, but higher latency as there will be a need to transfer data offsite for processing.
In this regard it is critical for latency sensitive applications to allow deployment on mobile (low-power) devices. 
To minimize latency (including wait-time), in addition to parallel deployment, the reconstruction algorithm should have a predictable and constant inference time, which is not necessarily true for iterative approaches.

Unfortunately, while some papers reported their training and inference times, (Table~\ref{Surveyed_papers}-Columns "Metrics-IS") it is not obvious to compare their time complexity given the variability in datasets, sampling patterns, hardware, and DL frameworks.
Overall, the offline training of DL methods bypasses the laborious online optimization procedure of conventional methods, and has the advantage of lower inference time over all but the simplest analytical.


\section{Conclusion, Discussion and Future Direction}
\label{depth:Conclusion}
Literature shows that DL-based image reconstruction methods have gained popularity over the last few years and demonstrated image quality improvements when compared to conventional image reconstruction techniques especially in the presence of noisy and limited data representation. 
DL-based methods address the noise sensitivity and incompleteness of analytical methods and the computational inefficiency of iterative methods.

\subsection{Discussion}

\hspace{15pt}\textbf{Learning:}
Unlike conventional approaches that work on a single image in isolation and require prior knowledge, DL-based reconstruction methods leverage the availability of large-scale datasets and the expressive power of DL to impose implicit constraints on the reconstruction problem.
While DL-based approaches do not require prior knowledge, their performance can improve with it.
By not being dependent on prior knowledge, DL-based methods are more decoupled from a specific imaging (sub)modality and thus can be more generalizable.
Real-time reconstruction is offered by DL-based methods by performing the optimization or learning stage offline, unlike conventional algorithms that require iterative optimization for each new image. The diagnostician can thus shorten diagnosis time increasing the throughput in patients treated.
In operating theatres and emergency settings this advantage can be life saving.

\textbf{Interpretability:}
While the theoretical understanding and the interpretability of conventional reconstruction methods are well established and strong (e.g., one can prove a method's optimality for a certain problem class), it is weak for the DL-based methods (due to the black-box nature of DL) despite the effort in explaining the operation of DL-based methods on many imaging processing tasks. However, one may accept the possibility that interpretabilty is secondary to performance as fully understanding DL-based approaches may never become practical.


\textbf{Complexity:}
On the one hand, conventional methods can be straightforward to implement, albeit not necessarily to design. 
On the other, they are often dependent on parameters requiring manual intervention for optimal results.
DL-based approaches can be challenging to train with a large if not intractable hyper-parameter space (\eg learning rate, initialization, network design). In both cases, the hyper-parameters are critical to results and require a large time investment from the developer and the practitioner.
In conclusion, there is a clear need for robust self-tuning algorithms, for both DL-based and conventional methods.

\textbf{Robustness:}
Conventional methods can provide good reconstruction quality when the measured signal is complete and the noise level is low, their results are consistent across datasets and degrade as the data representation and/or the signal to noise ratio is reduced by showing noise or artifacts (\eg streaks, aliasing). However, a slight change in the imaging parameters (\eg noise level, body part, signal distribution, adversarial examples, and noise) can severely reduce the DL-based approaches' performances and might lead to the appearance of structures that are not supported by the measurements~\citep{antun2019instabilities,gottschling2020troublesome}. DL based approaches still leave many unsolved technical challenges regarding their convergence and stability that in turn raise questions about their reliability as a clinical tool.
A careful fusion between DL-based and conventional approaches can help mitigate these issues and achieve the performance and robustness required for biomedical imaging.

\textbf{Speed:}
DL-based methods have the advantage in processing time over all but the most simple analytical methods at inference time. As a result, latency will be low for DL-based methods. However, one must be careful in this analysis. DL-based methods achieve fast inference by training for a long duration, up to weeks, during development. If any changes to the method are needed and retraining is required, even partial, a significant downtime can ensue. Typical DL-based methods are not designed to be adjusted at inference time. Furthermore, when a practitioner discovers that, at diagnosis time, the end result is sub-par, an iterative method can be tuned by changing its hyper-parameters. For a DL-based approach, this is non-trivial if not infeasible.

A final if not less important distinction is adaptive convergence. At deployment, a DL-based method has a fixed architecture and weights with a deterministic output. Iterative methods can be run iterations until acceptable performance is achieved. This is a double-edged sword as convergence is not always guaranteed and the practitioner might not know exactly how many more iterations are needed.

\textbf{Training Dataset:}
Finally, the lack of large scale biomedical image datasets for training due to privacy, legal, and intellectual property related concerns, limits the application of DL-based methods on health care. Training DL-based models often require scalable high performance hardware provided by cloud based offerings. However, deploying on cloud computing and transmitting the training data risks the security, authenticity, and privacy of that data.
Training on encrypted data offers a way to ensure privacy during training~\cite{gilad2016cryptonets}. More formally a homomorphic encryption algorithm~\cite{rivest1978data} can ensure evaluation (reconstruction) on the encrypted data results
are identical after decryption to reconstruction on the non-encrypted data. In practice, this results in an increase in dataset size as compression becomes less effective, a performance penalty is induced by the encryption and decryption routines, and interpretability and debugging the learning algorithm becomes more complex since it operates on human unreadable data.

The concept of federated learning, where improvements of a model (weights) are shared between distributed clients without having to share datasets, has seen initial success in ensuring privacy while enabling improvements in quality~\cite{geyer2017differentially}.
However a recent work~\cite{zhu2019deep} has shown that if an attacker has access to the network architecture and the shared weights, the training data can be reconstructed with high fidelity from the gradients alone. Data sharing security in a federated setting still presents a concern that requires further investigations.

Simulating a suitable training set also remains a challenge that requires careful tuning and more realistic physical models to improve DL-based algorithm generalization.

\subsection{Future Directions}
The future of the field will be to produce higher quality images given the most limited resource budget such as radiation dose, scanning time, and scanner complexity as well as from online real patient inputs using algorithms with the fewest hand-tuned parameters and lowest power consumption.

There remain a vast range of challenges and opportunities in the field.
So far, most approaches focus on CT and MR image reconstruction while only a handful of approaches exist for the reconstruction of the remaining modalities. Hence, the applicability of deep learning models to these problems is yet to be fully explored.
In addition, proposed deep learning architectures are often generic and are not fully optimized for specific applications. For instance, how to optimally exploit spatio-temporal redundancy, or how to exploit multi-spectral data.
By addressing these core questions and designing network layers to efficiently learn such data representation, the network architecture can gain a boost in performance and reliability.

Joint multi-modal image reconstruction, such as DOT/CT~\citep{Baikejiang2017} and PET/ MRI~\citep{Wang2015}, has been proposed on several iterative approaches to take advantage of both imaging modalities and led to improving the overall imaging performance, especially avoiding the spatio-temporal artifacts due to the scanning with different devices at different times and positions. The idea relies on leveraging the information provided by feature similarity between multiple modalities. While this direction has great interest, it has only just begun to receive consideration~\citep{cui2019ct} and remains a direction for further exploration. Collecting suitably calibrated and registered data on hybrid multi-modality imaging systems remains a key challenge as well.

Attention driven sampling received increasing interest recently especially in a limited data representation context. While adapting the sampling to the reconstruction algorithm showed improved image quality compared to conventional sampling strategies~\citep{jin2019self}, it could be computationally more expensive with unknown convergence behavior. The development of efficient sampling learning algorithms would be a promising research direction. In optical coherence tomography the application of wavelet based compressive sensing has been shown capable to reconstruct with a little as 20\% of the samples~\citep{lebed2013rapid}. The potential demonstrated through handcrafted sampling strategies indicates that DL-based methods could also exploit this opportunity.

Deep learning models require computationally powerful machines (GPU) to provide online reconstruction and achieve promised performance. Network pruning and sparsifying, recently proposed for computer vision tasks~\citep{Alford2018,Aghasi2017,Huang2018}, present a promising direction yet to be explored on image reconstruction tasks in order to allow DL-based model processing on CPU and mobile devices. This will be of great interest to emergent mobile scanners~(\eg DOT~\citep{ shokoufi2016}, US~\citep{2017portablexray},CT ~\citep{rykkje2019hand}).

Reducing the encoding size in bits of network weights without altering the quality of the prediction has been demonstrated on several references deep learning image processing networks~\cite{sun2019hybrid}. A reported decrease in training time of 30-60\% is only one benefit. By reducing network weights from 64 or 32 bits to 8 or smaller (\ie weights quantization) the network requires a smaller memory footprint. As a result, networks that are at the time of writing too large to fit on a single GPU can be reduced to fit on a single GPU. Conversely, networks too large to deploy on edge devices (handheld scanner, mobile phones, etc.) can become easily deployable in the field without changing their architecture. Finally and not least important, the reduction in training time results in a significant reduction in the environmental impact of the training procedure.


Finally, as performance evaluation metrics are more computer vision task-oriented, they tend to be insufficient in the biomedical imaging context as they do not provide real diagnostic accuracy significance. We advocate a shift toward task-oriented evaluations given that reconstructed images are usually used for a specific diagnosis or treatment purpose. While such a measure may be expensive, especially if they require human experts' feedback, they will be critical in creating algorithms that can advance the biomedical imaging practice.

\begin{landscape}
\clearpage
\onecolumn
\newgeometry{bottom=2cm,top=2cm}
\begin{table*}[t]

\begin{landscape}
\caption{Comparison between different papers that have applied machine learning to the biomedical image reconstruction problem. Mod.: modality; Samp.: signal sampling (F: full sampling, SS: sub-sampling, LA: limited angle data, SV: sparse viev, LD: low dose data); MS: multi-spectral input signal; TA: consideration of temporal aspect; D: dimension (PC: point clouds); Site: type of tissue (soft tissue like brain and breast, mix: different part of the body, Abdom: abdominal area, Bio Samp: biological samples, Sim: simulated data); Arch: architecture model; E2E: whether the approach is end-to-end or combined (Pos: DL as post-processing, Pre: DL as pre-processing, R2T: Raw-to-task); Loss: used loss function (Comb.loss: combined loss, Perc: perceptual loss, Adver: adversarial loss, CE: cross entropy, Jacc: Jaccard loss, MMD: maximum mean discrepancy); Reg: regularization; Inp: input data (M.Spec Meas: multi-spectral measurements, RF: radio frequencies measurements, PA Meas: photo-acoustic measurements); Out: output data (Chrom Maps: chromophore concentration map, Ref index: refractive index, Img: intensity image, Seg: segmentation mask); Metrics: evaluation metrics (M: MSE or RMSE, P: PSNR, S: SSIM, D: lesion detection metric like Dice and Jaccard, C: contrast to noise ratio and IT: reconstruction inference speed) Aug: used augmentation technique (A: affine, E: elastic, R: rotation transformations); Data: used training data (Sim: simulated data, Clini: clinical data, Phan: phantom data) Pub. Data: public data used.}
\label{Surveyed_papers}   
\begin{scriptsize}
\begin{longtable}[t]
{cc@{\hskip6pt} cc@{\hskip6pt} cc@{\hskip6pt} cc@{\hskip6pt} cc@{\hskip6pt} cc@{\hskip6pt} cc@{\hskip6pt} cc@{\hskip6pt} cc@{\hskip6pt} cc@{\hskip6pt} cc@{\hskip6pt} cc@{\hskip6pt} cc@{\hskip6pt} cc@{\hskip6pt} cc@{\hskip6pt} cc@{\hskip6pt} cc@{\hskip6pt} c}

     \\
\toprule
Ref
    &   \makecell[b]{Mod.}
        &   \makecell[b]{Year}    &   Samp.
            &   \makecell[b]{MS}
                &   \makecell[b]{TA}
                    &   \makecell[b]{D}  
                       &   \makecell[b]{Site} 
                           &   \makecell[b]{Arch.}
                               &   \makecell[b]{E2E}
                                   &   \makecell[b]{Loss}
                                      &   \makecell[b]{Reg.}
                                         &   \makecell[b]{Inp.}
                                            &   \makecell[b]{Out.}
                                              &   \makecell[b]{Aug.}
                                                 &   \makecell[b]{Metrics\\ M$|$P$|$S$|$D$|$C$|$IS}
                                                   &   \makecell[b]{Data}
                                                       &   \makecell[b]{Size}
                                                          &   \makecell[b]{Pub.\\Data} \\
    \midrule
\endfirsthead
    \toprule
~\citeauthor{cai2018end} & PA &2018&F& \checkmark& &2D& Mix&ResU-Net& \checkmark &MSE&  &\begin{tabular}[c]{@{}l@{}}M.Spec \\Meas\end{tabular} &\begin{tabular}[c]{@{}l@{}}Chrom\\ Maps\end{tabular} & &\checkmark$|$\checkmark$|$ -$|$ -$|$ -$|$\checkmark &Sim & $\sim$2000& \\ \addlinespace[3pt]

~\citeauthor{yoon2018efficient}&US &2019& SS & & &3D&Abdom&Decoder & \checkmark & MSE &L2 &RF&Img& &  -$|$\checkmark$|$ -$|$ -$|$ -$|$\checkmark&Clini&500 &    \\ 
\addlinespace[3pt] 

~\citeauthor{wu2018direct} & US & 2018 & F &  & \checkmark & 2D & Mix & CNN & \checkmark & MAE &  & RF & \begin{tabular}[c]{@{}l@{}}Elast\\dist\end{tabular}&  &\checkmark$|$\checkmark$|$ -$|$ -$|$\checkmark$|$ - &\begin{tabular}[c]{@{}l@{}}Sim\\ Phan\\ Clini\end{tabular}  & $\sim$100 & \\ \addlinespace[3pt]

~\citeauthor{gupta2018cnn}&CT&2018&SV& &  & 2D & Mix & CNN &Pos& \begin{tabular}[c]{@{}l@{}}Comb.\\loss\end{tabular} &  & FBP &Img &  & -$|$\checkmark$|$\checkmark$|$-$|$-$|$-& Clini& $\sim$1000&LowDoseCT\\ \addlinespace[3pt]

~\citeauthor{sun2018efficient}& \begin{tabular}[c]{@{}l@{}}Mult.\\Scat\end{tabular} &2018 & F &  &  & 2D &  \begin{tabular}[c]{@{}l@{}}Bio\\ Samp.\end{tabular} & ResU-Net & Pos & MSE &  & FBP &\begin{tabular}[c]{@{}l@{}}Ref\\index\end{tabular}&  & -$|$\checkmark$|$ -$|$ -$|$ -$|$ -&\begin{tabular}[c]{@{}l@{}} Sim \\ Phan\end{tabular} & 1550 &Fresnel\\ \addlinespace[3pt]

~\citeauthor{zhu2018image} & MRI & 2018 & SS &  & & 2D &Brain & Decoder & \checkmark & MSE & L2 & k-space &Img & A & \checkmark$|$\checkmark$|$ -$|$ -$|$ -$|$ - & \begin{tabular}[c]{@{}l@{}}ImagNet\\Clini\end{tabular}& $\sim$10K &MGH-USCHCP\\ \addlinespace[3pt]

~\citeauthor{chen2017lowb} & CT & 2017 & LD &  &  & 2D & Mix & Residual AE  &Pos & MSE &  & FBP& Img& A+R &\checkmark$|$\checkmark$|$\checkmark$|$ -$|$ -$|$ - &\begin{tabular}[c]{@{}l@{}}Sim \\ Clini\end{tabular}& $\sim$10K &LowDoseCT\\ \addlinespace[3pt]

~\citeauthor{wurfl2018deep} & CT & 2018 & LA &  &  & 3D & Mix &  \begin{tabular}[c]{@{}l@{}}Unrolling\\Iterative\end{tabular}& \checkmark & MSE &  & Sinogram & Img &   & -$|$\checkmark$|$\checkmark$|$ -$|$ -$|$ - & Clini & $\sim$10K  &LowDoseCT\\ \addlinespace[3pt]

~\citeauthor{feng2018back} & DOT & 2019 & F &  &  & 2D & Soft & MLP & \checkmark & MSE &  &Scattering& \begin{tabular}[c]{@{}l@{}}Chrom\\ Maps\\\end{tabular}&  &\checkmark$|$\checkmark$|$\checkmark$|$ -$|$ -$|$ - & Sim & $\sim$20K &  \\ \addlinespace[3pt]

~\citeauthor{yoo2019} & DOT & 2019 & F &  &  & 3D & Mix &Decoder & \checkmark & MSE &  & Scattering& \begin{tabular}[c]{@{}l@{}}Chrom\\ Maps\end{tabular}&  & -$|$ -$|$ -$|$ -$|$ -$|$ - & \begin{tabular}[c]{@{}l@{}} Sim\\Clini \end{tabular}& $\sim$1500  &  \\ \addlinespace[3pt]

~\citeauthor{cui2019ct}& PET & 2019 & F &  &  & 3D & Mix & 3DU-Net &Pos& MSE &  &  Sinogram & Img&  & -$|$ -$|$ -$|$ -$|$\checkmark$|$ - & \begin{tabular}[c]{@{}l@{}}Sim \\Clini \end{tabular}& 30 &  & \\ \addlinespace[3pt]

~\citeauthor{hyun2018deep} & MRI & 2018 & SS &  &  & 2D & Brain & U-Net &  Pre & MSE &  & k-space &\begin{tabular}[c]{@{}l@{}} Full \\k-space\end{tabular}&  & \checkmark$|$ -$|$ \checkmark$|$ -$|$ -$|$ - & Clini & 1400 & \\ \addlinespace[3pt]

~\citeauthor{waibel2018reconstruction} & PA & 2018 &LV  &  &  & 2D & Vessels & U-Net &\checkmark & MAE &  & \begin{tabular}[c]{@{}l@{}}PA\\ Meas\end{tabular} & Maps & R &\checkmark$|$ -$|$ -$|$ -$|$ -$|$ - & Sim & 3600 & \\ \addlinespace[3pt]

~\citeauthor{cheng2018highly} & MRI & 2018 &SS&  &  & 2D & Soft & CNN &Pre & MSE &  & k-space  & Img &  & \checkmark$|$\checkmark$|$\checkmark$|$ -$|$ -$|$\checkmark & Clini &65K & \\ \addlinespace[3pt]

~\citeauthor{liang2018improve} & CT & 2018 &SV&  &  & 2D & Mix & ResNet & Pre & MSE &  & \begin{tabular}[c]{@{}l@{}}Sparse\\ Sinogram\end{tabular} & \begin{tabular}[c]{@{}l@{}}\\ Sinog.\end{tabular}&  &\checkmark$|$ -$|$ -$|$ -$|$ -$|$ -& Clini & 17K& giveascan\\ \addlinespace[3pt]

\end{longtable}
\end{scriptsize}
\end{landscape}
\end{table*}
\clearpage
\begin{table*}[t]
\begin{landscape}
\begin{scriptsize}
\begin{longtable}[t]
{cc@{\hskip6pt} cc@{\hskip6pt} cc@{\hskip6pt} cc@{\hskip6pt} cc@{\hskip6pt} cc@{\hskip6pt} cc@{\hskip6pt} cc@{\hskip6pt} cc@{\hskip6pt} cc@{\hskip6pt} cc@{\hskip6pt} cc@{\hskip6pt} cc@{\hskip6pt} cc@{\hskip6pt} cc@{\hskip6pt} cc@{\hskip6pt} cc@{\hskip6pt} c}
\toprule
Ref
    &   \makecell[b]{Mod.}
        &   \makecell[b]{Year}    &   Samp.
            &   \makecell[b]{MS}
                &   \makecell[b]{TA}
                    &   \makecell[b]{D}  
                       &   \makecell[b]{Site} 
                           &   \makecell[b]{Arch.}
                               &   \makecell[b]{E2E}
                                   &   \makecell[b]{Loss}
                                      &   \makecell[b]{Reg.}
                                         &   \makecell[b]{Inp.}
                                            &   \makecell[b]{Out.}
                                              &   \makecell[b]{Aug.}
                                                 &   \makecell[b]{Metrics\\ M$|$P$|$S$|$D$|$C$|$IS}
                                                   &   \makecell[b]{Data}
                                                       &   \makecell[b]{Size}
                                                          &   \makecell[b]{Pub.\\Data} \\
    \midrule
\endfirsthead
    \toprule
~\citeauthor{jin2017deep} & CT & 2017 &SV&  &  & 2D & Mix & U-Net & Pos & MSE &  & FBP &  Img & A & -$|$\checkmark$|$ -$|$ -$|$ -$|$ -  &\begin{tabular}[c]{@{}l@{}}Sim\\Clini\end{tabular}& 1000 & \\ \addlinespace[3pt]

~\citeauthor{wolterink2017generative} & CT & 2017 &LD&  &  & 2D & \begin{tabular}[c]{@{}l@{}}Thorax \\cardiac\end{tabular} & GAN & Pos & Adver& L2 &FBP& Img &  &\checkmark$|$\checkmark$|$ -$|$ -$|$ -$|$\checkmark & \begin{tabular}[c]{@{}l@{}}Sim\\Clini\end{tabular}&500 & \\ \addlinespace[3pt]

~\citeauthor{yang2017dagan} & MRI & 2017 & SS &  &  & 2D & Brain & \begin{tabular}[c]{@{}l@{}}Condi\\ GAN.\end{tabular}& Pos & \begin{tabular}[c]{@{}l@{}}Adver\\ Perc\end{tabular} &  & k-space & Img &A+E&\checkmark$|$\checkmark$|$ -$|$ -$|$ -$|$\checkmark & Clini &21K &MICCAI2013\\ \addlinespace[3pt]

~\citeauthor{qin2018convolutional} & MRI & 2019 &SS &  &\checkmark& 3D+t & Cardiac & RCNN & \checkmark & MSE &  & k-space & Img &\begin{tabular}[c]{@{}l@{}}A+E\\+R\end{tabular}& \checkmark$|$\checkmark$|$\checkmark$|$ -$|$ -$|$\checkmark&\begin{tabular}[c]{@{}l@{}}Sim\\Clini\end{tabular} & NA &  \\ \addlinespace[3pt]

~\citeauthor{sun2016deep} & MRI & 2016 &SS &  &  & 2D & \begin{tabular}[c]{@{}l@{}}Brain \\Chest\end{tabular} & \begin{tabular}[c]{@{}l@{}}Unrolling\\  ADMM\end{tabular} & \checkmark & NMSE &  & k-space &Img &  &\checkmark$|$\checkmark$|$ -$|$ -$|$ -$|$\checkmark& Clini & 150 &MICCAI2013\\ \addlinespace[3pt]

~\citeauthor{Schlemper2018} & MRI & 2018 & SS&  & \checkmark & 2D & Cardiac & CNNs& \checkmark & MSE &  & k-space & Img &A+E &\checkmark$|$\checkmark$|$ -$|$ -$|$ -$|$\checkmark& Sim & 300 & \\ \addlinespace[3pt]

~\citeauthor{nehme2018deep}& SMLM & 2018 &  &  &  &  2D& Sim &\begin{tabular}[c]{@{}l@{}}Encoder\\Decoder\end{tabular}& \checkmark & MSE & L1 &\begin{tabular}[c]{@{}l@{}}Fluophore\\emission\end{tabular} &\begin{tabular}[c]{@{}l@{}}SR\\Img\end{tabular}&  & \checkmark$|$ -$|$ -$|$ -$|$ -$|$\checkmark  &  Sim &  100K &SMLM\\ \addlinespace[3pt]

~\citeauthor{schlemper2017deep} & MRI & 2017 &  SS &  &  & 2D &Cardiac & CNNs & \checkmark & MSE &  & k-space & Img & A &\checkmark$|$ -$|$ -$|$ -$|$ -$|$\checkmark& Sim & 300& \\ \addlinespace[3pt]

~\citeauthor{boublil2015spatially} & CT & 2015 & LD&  &  & 2D &Mix & MLP & Pos & MSE &  & FBP & img &  &\checkmark$|$\checkmark$|$\checkmark$|$ -$|$ -$|$ -& Clini & 100K & \\ \addlinespace[3pt]

~\citeauthor{thaler2018sparse} & CT & 2018 &SV&  &  & 2D & Mix & wGAN & \checkmark & Adver &  &\begin{tabular}[c]{@{}l@{}}LD\\ Sinogram\end{tabular} &Img &R+A&\checkmark$|$ -$|$\checkmark$|$ -$|$ -$|$ -& Clini & NA & \\ \addlinespace[3pt]

~\citeauthor{zhou2019limited} & CT & 2019 &LA&  &  & 3D & Mix & GAN & \checkmark & \begin{tabular}[c]{@{}l@{}}MSE\\Adver.\end{tabular} &\begin{tabular}[c]{@{}l@{}}Cor\\coef\end{tabular} &\begin{tabular}[c]{@{}l@{}}LD\\Sinogram\\ FBP\end{tabular}&Img&  &\checkmark$|$ -$|$\checkmark$|$ -$|$ -$|$\checkmark& Clini & 5436 &LowDoseCT\\ \addlinespace[3pt]

~\citeauthor{yedder2018deep} & DOT & 2018 & LA &  &  & 2D & Breast & \begin{tabular}[c]{@{}l@{}}Decoder\end{tabular} &\checkmark & MSE & L2 &Scattering& \begin{tabular}[c]{@{}l@{}}Chrom\\ Maps\end{tabular}&  &\checkmark$|$\checkmark$|$\checkmark$|$\checkmark$|$ -$|$\checkmark & Phan &4000  & \\ \addlinespace[3pt]

~\citeauthor{wang2016accelerating} & MRI & 2016 &SS &  &  & 2D & Brain & Decoder & Pos& MSE &  & IFT & Img &  &  -$|$ -$|$ -$|$ -$|$ -$|$ - & Clini & 500 & \\ \addlinespace[3pt]


~\citeauthor{Sun2019} & MRI & 2019 & SS &  &  & 2D & Brain & Enc-Dec & R2T& \begin{tabular}[c]{@{}l@{}}MSE\\CE\end{tabular} &  & IFT &\begin{tabular}[c]{@{}l@{}}Img\\seg\end{tabular}&A+E & \checkmark$|$\checkmark$|$ -$|$\checkmark$|$ -$|$ - & Clini & 192  & MRBrainS\\ \addlinespace[3pt]

~\citeauthor{yedder2019limited} & DOT & 2019 & LA &  &  & 2D & Breast & \begin{tabular}[c]{@{}l@{}}Decoder\end{tabular} &\checkmark & \begin{tabular}[c]{@{}l@{}}MSE\\ Jacc.\end{tabular} & L2 &Scattering& \begin{tabular}[c]{@{}l@{}}Chrom\\ Maps\end{tabular}&  &\checkmark$|$\checkmark$|$\checkmark$|$\checkmark$|$ -$|$\checkmark & Phan &20K  & \\ \addlinespace[3pt]

~\citeauthor{boyd2018deeploco} & SMLM & 2018 & SS &  &  & PC & NA & CNN &\checkmark& MMD&  & \begin{tabular}[c]{@{}l@{}}Fluophore\\emission\end{tabular} &\begin{tabular}[c]{@{}l@{}}3D\\Localiz\end{tabular}&A & \checkmark$|$ -$|$ -$|$\checkmark$|$\checkmark$|$ - & Sim & 100K& SMLM\\ \addlinespace[3pt]

~\citeauthor{haggstrom2019deeppet} & PET & 2019 & F &  &  & 2D &Sim& \begin{tabular}[c]{@{}l@{}}Encoder\\Decoder\end{tabular}&\checkmark & MSE&  &Sinogram& Img & A+R & \checkmark$|$\checkmark$|$\checkmark$|$ -$|$ -$|$\checkmark& Sim &291K &  \\ \addlinespace[3pt]

~\citeauthor{oksuz2019detection}& MRI & 2019 & SS &  & \checkmark &2D+t& Cardiac &\begin{tabular}[c]{@{}l@{}}3D CNN\\RCNN\end{tabular} &\checkmark &\begin{tabular}[c]{@{}l@{}}MSE\\CE\end{tabular}&  & k-space & Img &  & \checkmark$|$\checkmark$|$\checkmark$|$ -$|$ -$|$ - &   \begin{tabular}[c]{@{}l@{}}Sim\\Clini\end{tabular}& 300&mridata.org\\ \addlinespace[3pt]

~\citeauthor{rajagopal2019towards}& CT & 2019 &SS&  & &3D & Phan &\begin{tabular}[c]{@{}l@{}}Unrolling\\Iterative\end{tabular}&\checkmark& MSE & & Sinogram & Img & &\checkmark$|$ -$|$ -$|$ -$|$ -$|$ - & Phan &50K& \\ \addlinespace[3pt]

~\citeauthor{huang2019data}& CT & 2019 &LA & & &2D & abdomen&\begin{tabular}[c]{@{}l@{}}U-Net\\Iterative\end{tabular}&\checkmark& MSE & & FBP & \begin{tabular}[c]{@{}l@{}}Artifact\\Img\end{tabular}& &\checkmark$|$ -$|$ -$|$ -$|$ -$|$ - & Clini &500& \\ \addlinespace[3pt]

~\citeauthor{wang2019optimization}& DOT & 2019 & LA &  &  & 2D & Phan &Stacked AE &\checkmark & MSE & &Scattering& \begin{tabular}[c]{@{}l@{}}Chrom\\ Maps\end{tabular}&  &-$|$-$|$-$|$\checkmark$|$ -$|$- & Sim &NA  & \\ 

\end{longtable}
\end{scriptsize}
\end{landscape}
\end{table*}
\end{landscape}
\normalsize
\restoregeometry 

\noindent \textbf{Acknowledgments} We thank the Natural Sciences and Engineering Research Council of Canada (NSERC) for partial funding. 

{\small
\bibliographystyle{abbrvnat}
\bibliography{main}
}

\end{document}